# DELAY PERFORMANCE OPTIMIZATION FOR MULTIUSER DIVERSITY SYSTEMS WITH BURSTY-TRAFFIC AND HETEROGENEOUS WIRELESS LINKS

*Jalil S. Harsini, and Farshad Lahouti*

WMC Lab., School of Electrical and Computer Engineering, University of Tehran

**ABSTRACT**

This paper presents a cross-layer approach for optimizing the delay performance of a multiuser diversity system with heterogeneous block-fading channels and a delay-sensitive bursty-traffic. We consider the downlink of a time-slotted multiuser system employing opportunistic scheduling with fair performance at the medium access (MAC) layer and adaptive modulation and coding (AMC) with power control at the physical layer. Assuming individual user buffers which temporarily store the arrival traffic of users at the MAC layer, we first present a large deviations based statistical model to evaluate the delay-bound violation of packets in the user buffers. Aiming at minimizing the delay probability of the individual users, we then optimize the AMC and power control module subject to a target packet-error rate constraint. In the case of a quantized feedback channel, we also present a constant-power AMC based opportunistic scheduling scheme. Numerical and simulation results are provided to evaluate the delay performance of the proposed adaptation schemes in a multiuser setup.

## 1. INTRODUCTION

In the context of wireless data networks, the increasing demand for high-speed multimedia services with stringent quality of service (QoS) requirements, highly motivates the development of cross-layer design paradigms for efficient utilization of the scarce radio resources [1]. However, due to the time-varying nature of wireless channels and also the stochastic characteristic of the arrival traffic, providing an efficient and low-complexity integrated design for such networks is challenging. In this direction, the current work presents a cross-layer framework for minimizing the queuing delay performance of a multiuser system employing opportunistic scheduling with fair performance at the medium access (MAC) layer and adaptive transmission at the physical layer.

It is well known that link adaptation based on adaptive modulation and coding (AMC) and power control is a powerful technique for improving the QoS performance over wireless channels with time-varying fading [2][3]. In addition, in multiuser wireless systems, where users channels suffer from independent time-varying fading, a channel-quality based opportunistic scheduler can improve the throughput performance by exploiting the multiuser diversity [4]. In order to increase the spectral efficiency and also to enable exploiting multiuser diversity, adaptive transmission is integrated with opportunistic scheduling in [5]-[8]. Aiming at reducing the feedback load of multiuser diversity systems, in [5] a joint opportunistic scheduler with constant-power discrete-rate adaptive modulation is analyzed.

Considering homogenous independent and identically distributed (i.i.d.) users channels, in [6] and [7] power and rate adaptation policies are presented to maximize the spectral efficiency of an adaptive modulation based opportunistic scheduling scheme while guaranteeing a required bit-error-rate performance. However, the issues related to link level queuing are not addressed. Also in [8], a vacation queuing model is adopted to analyze the (queuing) delay performance of a constant-power AMC based multiuser diversity system with Bernoulli traffic.

In wireless packet networks with real-time multimedia services, data arrivals have a random and bursty nature, resulting in a queuing delay. For these services delay-bound guarantee is a vital QoS performance measure. As a result, in such systems, establishing a cross-layer framework for optimizing the packet delay performance is of great importance. This is the main focus of the current paper. We consider a multiuser diversity system with heterogeneous wireless links employing proportional fair opportunistic scheduling based on relative signal-to-noise ratio (SNR) [4]. Assuming a bursty and delay-sensitive packet traffic for the individual users, we present an analytical framework for link adaptation design including AMC and power control, which minimizes the delay-bound violation probability for users packets at the MAC layer. In order to model the delay-bound violation, we use a large deviations based statistical description. In particular, we optimize the AMC and power control scheme at the physical layer subject to a target packet-error rate (PER) and an average transmit power constraint. The proposed adaptation policy depends not only on the channel conditions, but also on the buffer backlog and arrival traffic characteristics in a cross-layer manner. In case, where only the quantized values of the users SNRs are fed back to the BS, we also present a constant-power AMC based fair opportunistic scheduling scheme. Numerical results are provided to illustrate the delay performance of the proposed adaptation schemes under different conditions of the traffic source and users distributions.

## 2. SYSTEM MODEL

We consider the downlink of a multiuser packet communication system in which a base station (BS) serves $K$ users using time-division multiplexing (Fig. 1). The packets generated by the traffic sources are first stored in $K$ transmit buffers at the BS, each corresponding to an individual user. By dividing the time into fixed-length slots a scheduler at the BS selects a user for transmission in each time-slot based on the channel state information (CSI) provided by the users through feedback channels. The packets of the selected user are then grouped into a frame structure and are transmitted at the physical layer, where AMC and power adaptation are employed to enhance the delay performance. We consider the packet and frame structures as follows. Each packet contains a fixed number of bits, and a data frame which has a fixed time duration equal to one time-slot, includes a fixed number of symbols. Since the number of symbols which represent a packet depends on the employed AMC mode, each frame carries a variable number of packets. Upon reception of the transmitted data frame by the selected user, the bit streams are first decoded and then are placed into a packet structure. The following assumptions are made: (1) a strict delay bound for packets is considered, i.e., if a packet is not transmitted before a deadline, it is dropped from the corresponding user queue and a packet loss is declared; (2) a strong cyclic-redundancy-check code is used so that a perfect packet error detection is possible; (3) in case a packet is detected erroneous, it is considered lost; (4) the CSI of all users are fed back reliably to the BS without delay.

## 2.1. Channel Model and Transmission Scheme

We assume that the $i$th user experiences a frequency flat-fading channel with stationary and ergodic time-varying gain $h_i$, and additive white Gaussian noise (AWGN) with zero mean and variance $\sigma^2$. A block-fading channel model is adopted, where the channel gain $h_i$ remains constant during a frame length and is i.i.d. from one frame to another. This suits wireless links with slowly moving terminals [9] and is also adopted by [4] for analyzing multiuser diversity systems. Transmitting with the average power $\bar{S}$ to the user $i$ at the frame $n$, results in the received SNR $\gamma_n^{(i)} = \bar{S}.|h_i(n)|^2/\sigma^2$. By virtue of stationary assumption, the probability density function (PDF) of $\gamma_n^{(i)}$ is independent of $n$, and may be simply denoted by $f_{\gamma^{(i)}}(\gamma)$.

In this paper, we consider an independent but non-identically distributed (i.n.d.) channel environment, where different users have different average channel SNRs. For this scenario, a normalized SNR-based (NS) opportunistic scheduler, which is developed based on the idea of the original proportional fair scheduling, enhances the system throughput by exploiting the multiuser diversity and facilitates fair performance [4][5]. Let $\bar{\gamma}^{(i)}$ and $x^{(i)} = \gamma^{(i)}/\bar{\gamma}^{(i)}$ denote the average channel SNR and the normalized SNR for user $i$. Considering a specific fading distribution, e.g., Rayleigh, it is straightforward to show that all $x^{(i)}$ variables are independent with the same PDF, denoted by $f_{x^{(i)}}(x) = f_x(x), \forall i$ [5]. To employ joint NS opportunistic scheduling and adaptive transmission, it is assumed that the $i$th user feedbacks $x^{(i)}$ to the transmitter prior to transmission of each frame. The NS scheduler at the $n$th frame selects one user with the largest normalized SNR and assigns the current frame to the selected user for transmission.

The AMC and power control are employed as described below. We partition the total range of the normalized SNR into $M$ consecutive intervals, defined by the threshold points $\xi_0(=0) < \xi_1 < \xi_2 < \cdots < \xi_M(=\infty)$. Whenever the normalized SNR of the selected user falls within the interval $[\xi_{m-1}, \xi_m)$, $m = 1,2,...,M$, the mode $m$ of AMC is chosen, data is transmitted to the selected user with rate $R_m$ (bits/symbol) and power $S_m^{(i^*)}(x^*)$ (watts), where $i^*$ and $x^* = \max_i x^{(i)}$, denote the index and the normalized SNR of the scheduled user.

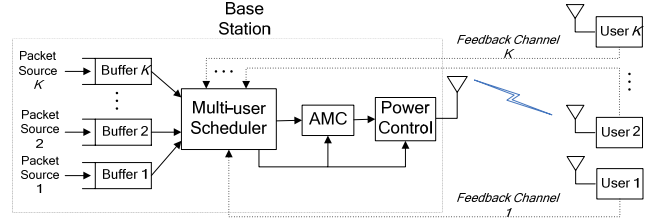

Figure 1. System model with multiuser traffic.

To simplify the analysis, in the presence of AWGN, we use the following expression to approximate PER in mode $m$ for $i$th user as a function of the post-adaptation SNR $\gamma^{(i)} S_m^{(i)}(x^{(i)})/\bar{S}$

$$PER_m^{(i)}(\gamma^{(i)}) \approx \frac{1}{1 + \left(a_m . \gamma^{(i)} S_m^{(i)}(x^{(i)})/\bar{S}\right)^{g_m}} \quad (1)$$

The parameters $\{a_m, g_m\}$ are mode and packet-size dependent constants and can be obtained by least square fitting the PER expression of (1) to the exact PER computed via simulations. This model is verified in section 6.

## 2.2. Traffic Source Model

In order to characterize the bursty nature of the buffer arrival traffic, we use a Markov modulated Poisson process (MMPP) model [10]. This model is suitable for characterizing various types of multimedia packet traffic such as voice, MPEG video and self-similar traffic. For more details see [10] and the references therein. In addition, we assume that all users have identical queue arrival statistics. This is motivated, e.g., in a cellular system where the BS serves multiple users with the same traffic service. As elaborated in the next section, in order to describe the delay probability we use the log moment generating functions for queue arrival and service processes, also referred to as the Gärtner-Ellis (GE) limit [11]. If $A(0,\tau)$ denotes the amount of bits generated by the source in the interval $(0,\tau)$, the GE limit of the arrival process is given by

$$\Lambda_A(s) = \lim_{\tau \to \infty} \frac{1}{\tau} \log[E\{\exp(s.A(0,\tau))\}]. \quad (2)$$

## 3. DESCRIPTION OF DELAY PROBABILITY

In this section, we introduce a large deviations based statistical model to evaluate the packet delay-bound violation probability in the individual user buffers. Since under NS scheduling all users have the same opportunity to access the shared channel (it is proved in section 4), the packet delay performance can be analyzed for an arbitrary user buffer. For convenience, in the followings the first user is selected as the arbitrary user, and we, hereafter, refer to it the tagged user.

### 3.1. Queue Service Process of the Tagged User

Here, we describe the GE limit of the queue service process for the tagged user. When AMC is employed, a variable number of packets are transmitted in each frame. Let us assume that the tagged user is selected by the scheduler to transmit its data in frame $n$. We denote the number of packets that can be accommodated per frame in the transmission mode 1, by $b$. The parameter $b$ may be related to the number of time-slots allocated to the assumed connection per frame. If $c_j$ denotes the number of packets transmitted per frame, in the AMC mode $j$, we have $c_j = bR_j/R_1, j \geq 1$. As a result, the queue service process for the tagged user, denoted by $\{C_n, n=1,2,...\}$, is a discrete-time stationary stochastic process with a total of $M+1$ service rates $\{c_j\}_{j=0}^{M}, (c_0 = 0)$, each with the probability $P_j$ given by

$$P_j = \begin{cases} \Pr\left(x^{(1)} \geq \tilde{x}, \; x^{(1)} \in [\xi_{j-1}, \xi_j]\right), & j \geq 1 \\ \Pr\left(x^{(1)} < \tilde{x}\right), \text{(tagged user is not scheduled)} & j = 0 \end{cases} \quad (3)$$

where $\tilde{x} = \max_{i, i \neq 1} x^{(i)}$. Let $C(0,\tau) = \sum_{n=1}^{\tau/T_f} N_b C_n$ (bits) represent the accumulative queue service process of the tagged user in the interval $[0,\tau]$, where $N_b$ is the number of bits in a packet, and $T_f$ is the frame length. Assume that the GE limit of $C(0,\tau)$, $\Lambda_C(s)$, exists and is finite and differentiable for all real $s$. Since we assume independent channel processes for different users and a block-fading model for each user channel, the rate sequence $\{C_n\}$ is an uncorrelated process; in this case from equation (3), we have

$$\Lambda_C(s) = \frac{1}{T_f} \log[E\{\exp(sN_b C_n)\}] = \frac{1}{T_f} \log\left[\sum_{j=0}^{M} P_j \cdot \exp(sN_b c_j)\right]. \quad (4)$$

The next Proposition presents the probability $P_j$ in equation (3).

*Proposition* 1: For an AMC-based NS scheduling scheme with $K$ users, the probability $P_j, j \geq 1$, in (3) is given by

$$P_j = \int_{\xi_{j-1}}^{\xi_j} f_x(x) dx \cdot \int_0^{\xi_{j-1}} f_{\tilde{x}}(x) dx + \int_{x=\xi_{j-1}}^{\xi_j} f_{\tilde{x}}(x) \int_{y=x}^{\xi_j} f_x(y) dy dx \quad (5)$$

where $f_{\tilde{x}}(x)$ denotes the PDF of the random variable (r.v.) $\tilde{x}$. The probability $P_0$ is also obtained as

$$P_0 = \int_0^{\infty} f_{\tilde{x}}(x) \int_{y=0}^{x} f_x(y) dy dx. \quad (6)$$

*Proof*: The proof is provided in Appendix 1.

### 3.2. Delay-Bound Violation Probability

Let $D$ be the delay experienced by a packet in the *tagged* queue at the steady state, and $D_{\max}$ denote the delay-bound. It is known that if there exists a unique positive $s^*$ such that

$$\Lambda_A(s^*) + \Lambda_C(-s^*) = 0 \quad (7)$$

then under appropriate conditions, the tail probability of the form $\Pr(D > D_{\max})$ for the tagged user can be approximated as [11][12]

$$P_d(D_{\max}) = \Pr(D > D_{\max}) \approx \Pr(D > 0) e^{\Lambda_C(-s^*)D_{\max}}. \quad (8)$$

The quantity $\Pr(D > 0)$ can be estimated as

$$\Pr(D > 0) = \sum_{j=0}^{M} P_j \cdot \Pr(D > 0 | C_n = c_j) \quad (9)$$

where the term $\Pr(D > 0 | C_n = c_j)$ can be estimated by empirical averaging using a number of source traffic samples [12]. In this work, we use (8) to evaluate the delay-bound violation probability in a queuing system, whose arrival packets have an expiration deadline for transmission. This model is verified in section 6.

## 4. OPTIMIZING THE DELAY PERFORMANCE

In this section, we propose a link adaptation scheme including AMC and power adaptation, such that the delay-bound violation probability for the tagged user is minimized subject to a target PER and an average transmit power constraint.

In order to minimize the delay probability in (8), we optimize a power adaptation and AMC scheme such that the *delay exponent* $\Lambda_C(-s^*)$ is minimized. Since the function $\Lambda_C(-s)$ (resp. $\Lambda_A(s)$) is negative (positive) and decreasing (increasing) for $s>0$ [3], given fixed source parameters, from equation (7) we see that if the function $\Lambda_C(-s)$ is minimized over each point $s$, then the exponent $\Lambda_C(-s^*)$ is also minimized. Since logarithm is a monotonically increasing function, from (4) minimizing $\Lambda_C(-s)$ is equivalent to

$$\min \Lambda_C(-s) \equiv \min \sum_{j=0}^{M} \theta_j(s) \cdot P_j(\xi_{j-1}, \xi_j) \quad (10)$$

where $\theta_j(s) = \exp(-sN_b c_j)$. Note that in (10), we use the notation $P_j(\xi_{j-1}, \xi_j)$, to emphasize that $P_j$ is a function of $\xi_{j-1}$ and $\xi_j$ as specified in (5). Consider the instantaneous PER constraint, so that $PER_m^{(i)}(\gamma^{(i)}) = P_{\text{tgt}}, \forall m,i$, where $P_{\text{tgt}}$ denotes the *target* PER. Using these constraints and (1), we find the following expression for power adaptation in mode $m$

$$\frac{S_m^{(i^*)}(x^*)}{\overline{S}} = \frac{d_m}{\overline{\gamma}^{(i^*)} x^*}, \quad \xi_{m-1} \leq x^* \leq \xi_m \quad (11)$$

in which $d_m = (1/a_m)\left(1/P_{\text{tgt}} - 1\right)^{(1/g_m)}$. Let $P_s^{(i)}$ denote the probability that the $i$th user is scheduled by the BS. Using equations (3) and (6), we obtain $P_s^{(i)} = 1 - P_0, \forall i$, which is independent from the user index $i$. This means that all users in the system have the same opportunity to be selected by the BS. Accordingly, the average transmit power is given by

$$E\{S_m^{(i^*)}(x^*)\} = \sum_{i=1}^{K} P_s^{(i)} E\{S_m^{(i^*)}(x^*) | i^* = i\}$$
$$= C_p \overline{S} \sum_{m=1}^{M} d_m \cdot \int_{\xi_{m-1}}^{\xi_m} \frac{1}{x} f_{x^*}(x) dx \quad (12)$$

where $C_p = (1-P_0) \sum_{i=1}^{K} 1/\overline{\gamma}^{(i)}$, and $f_{x^*}(x)$ denotes the PDF of the post scheduling normalized SNR $x^*$.

Now, we use the Lagrange method to minimize the packet delay probability of the tagged user. The Lagrange equation is

$$L(\xi_1, ..., \xi_{M-1}, \lambda) = \sum_{m=0}^{M} \theta_m(s) \cdot P_m(\xi_{m-1}, \xi_m)$$
$$+ \lambda \left(C_p \sum_{m=1}^{M} d_m \cdot \int_{\xi_{m-1}}^{\xi_m} \frac{1}{x} f_{x^*}(x) dx - 1\right) \quad (13)$$

where $\lambda > 0$ is a Lagrange multiplier and the last term in the RHS of eq. (13) represents the average transmit power constraint in the form of $E\{S_m^{(i^*)}(x^*)\} \leq \overline{S}$. From (13), we obtain

$$\frac{\partial L}{\partial \xi_m} = (\theta_m(s) - \theta_{m+1}(s)) f_x(\xi_m) \int_0^{\xi_m} f_{\tilde{x}}(x) dx$$
$$+ \lambda C_p (d_m - d_{m+1}) \frac{f_{x^*}(\xi_m)}{\xi_m}, \quad \forall m \geq 1 \quad (14)$$

Let $F_x(x)$ denote the cumulative distribution function (CDF) corresponding to the PDF $f_x(x)$. With $\tilde{x} = \max_{i, i \neq 1} x^{(i)}$ and $x^* = \max_i x^{(i)}$, it is straightforward to show that [4]

$$f_{\tilde{x}}(x) = (K-1) f_x(x) [F_x(x)]^{K-2}, \quad (15)$$
$$f_{x^*}(x) = K \cdot f_x(x) [F_x(x)]^{K-1}. \quad (16)$$

Substituting (15) and (16) into (14) and setting $\partial L/\partial \xi_m = 0$, after some manipulations we obtain the following AMC threshold points

$$\xi_m = \frac{KC_p(d_{m+1} - d_m)}{\theta_m(s) - \theta_{m+1}(s)}\lambda, \quad m = 1,...,M-1. \quad (17)$$

**Remark 1**: In order to determine the threshold points in (17), we need to obtain the constant $\lambda$ and $s^*$ in (7) by solving two nonlinear equations, i.e., equation (7) and average power constraint satisfied with equality, $E\{S_m^{(i^*)}(x^*)\} = \bar{S}$. To this end, we use a bisection root-finding method [13] to find the roots of these equations. Once the pair $(s^*, \lambda)$ is obtained, the adaptation solution is specified by (11) and (17).

**Remark 2**: In [14], a link adaptation scheme is presented for a single-user communication scenario which considers an outage mode to avoid a deep fading condition. As a result, the power control policy proposed in [11] uses a limited average transmit power to compensate the channel fading. In contrast, in our multiuser setup with $K > 1$, the probability that the transmitter encounters a deep fade condition is negligible. In fact, from (16) it is clear that the PDF $f_{x^*}(x)|_{x=0} = 0$. Therefore, provisioning of an outage mode is not necessary.

## 5. NS SCHEDULING WITH QUANTIZED FEEDBACK

When only the quantized CSI are fed back to the BS, a NS scheduling with a constant-power (CP) AMC is employed. In [5], a simple approach for CP-AMC based NS scheduling is presented, which partitions the normalized SNR range into equal probability SNR regions. However, such design can not in general guarantee a required target PER.

Here, we present a CP-AMC based NS scheduling scheme which explicitly guarantees a target PER. Since in this case the scheduling decision is not directly based on the largest normalized SNR fed back from users, we consider a transmission outage mode, so that no data is sent in this mode. Accordingly, to make a scheduling decision, we partition the normalized SNR range into $M+1$ consecutive intervals as $[\xi_m, \xi_{m+1}]$, $m = 0,1,2,...,M$, where $\xi_0 = 0$, $\xi_{M+1} = \infty$, and the interval $[\xi_0, \xi_1)$ corresponds to the outage mode. Let $J_n^{(i)}$ be the index of the normalized SNR level for the $i$th user in the $n$th frame, as follows

$$J_n^{(i)} = m, \quad \text{if} \quad \xi_m \leq x_n^{(i)} < \xi_{m+1}, \quad m = 1,...,M \quad (18)$$

It is assumed that the $i$th user feedbacks the index $J_n^{(i)}$ to the BS prior to transmission of the $n$th frame. The CP-AMC based NS scheduler at the frame $n$ selects one user randomly among those with the index $J_n^* = \max_i J_n^{(i)}$, and assigns the current frame to the selected user for transmission. Moreover, when the index $J_n^* = m$, the mode $m$ of AMC is chosen for transmission with a constant transmit power equal to $\bar{S}$. For this scheme, the next Proposition presents the probabilities $P_j$ in equation (4).

*Proposition* 2: For an AMC based NS scheduling scheme with quantized feedback, the probability $P_j, j \geq 1$, in (4) is given by

$$P_j = \Pr\left(J_n^{(1)} = J_n^* = j, V_n \leq 1/U_n\right)$$
$$= \frac{1}{K}\left([F_x(\xi_{j+1})]^K - [F_x(\xi_j)]^K\right) \quad (19)$$

Here, the r.v. $U_n$ represents the number of users with the index $J_n^*$ at the beginning of the frame $n$, and $V_n$ is an i.i.d. r.v. uniformly distributed on the interval $[0,1)$. The probability $P_0$ is given by

$$P_0 = 1 - \sum_{j=1}^{M} P_j = 1 - \frac{1}{K}\left(1 - [F_x(\xi_1)]^K\right). \quad (20)$$

*Proof*: The proof is provided in Appendix 2.

To guarantee the PER constraint $PER_m^{(i)}(\gamma^{(i)}) \leq P_{tgt}$, $\forall m \geq 1$, for the $i$th user we set the AMC threshold in mode $m$, denoted by $\xi_m^{(i)}$, to the minimum SNR required to achieve the PER $P_{tgt}$, i.e.,

$$\xi_m^{(i)} = d_m/\bar{\gamma}^{(i)}, \quad m = 1,2,3,...,M. \quad (21)$$

Obviously, the design in (21) depends on the scheduled user. In order to present an AMC scheme which satisfies the target PER constraint $P_{tgt}$ for all users, we use the following design

$$\xi_m = \max_i(\xi_m^{(i)}), \quad m = 1,2,...,M. \quad (22)$$

## 6. NUMERICAL AND SIMULATION RESULTS

In this section, we present numerical and simulations results to evaluate the delay performance of the tagged user.

We select an AMC scheme with $M=7$ modes adopted from the IEEE 802.11a standard. Each mode is constructed by a convolutionally coded $M_n$-QAM scheme. Table I presents the transmission modes of the AMC scheme and the set of PER fitting parameters obtained for packet length $N_b=1080$ bits.

For numerical results, we consider a Rayleigh fading model for the channel gains, i.e., the PDF of the channel SNR for the $i$th user is $f_{\gamma^{(i)}}(\gamma) = (1/\bar{\gamma}^{(i)})\exp(-\gamma/\bar{\gamma}^{(i)}), \gamma \geq 0$. In this case, we have $f_{x^{(i)}}(x) = f_x(x) = \exp(-x), x \geq 0; \forall i$. To simulate an i.n.d. fading environment with $K$ users, it is assumed that different users are located in $K$ different tiers inside a cell. We consider a scenario in which the mean of the average SNRs for each user set remains constant and is equal to 12 dB. Two user sets are selected: the first set has $K=5$ users with $\bar{\gamma}^{(i)} \in \{10,11,12,13,14\}$ (dB), and the second set has $K=9$ users with $\bar{\gamma}^{(i)} \in \{8,9,10,11,12,13,14,15,16\}$ (dB). We also set the frame duration $T_f = 2$ms, the target PER $P_{tgt}=0.01$, the bandwidth parameter $b=2$, and the delay-bound $D_{max}=60$ms ($=30T_f$). The arrival source is modeled as a two-state ON-OFF MMPP (refer to [3] for the GE limit). The default parameters for the Markov source are: mean ON-state period $T_{ON}=10T_f$, mean OFF-state period $T_{OFF}=10T_{ON}$. For this source, the average rate is given by $\rho_s = \nu/(1 + T_{OFF}/T_{ON})$, where $\nu$ denotes the Poisson average rate in the ON state.

Fig. 2 verifies the PER modeling in equation (1) based on the fitting parameters in Table I. This figure shows that the PER model in equation (1) well approximates the exact PER.

In order to validate the delay probability model in (8), we simulate a queuing system for the tagged user for both considered user sets. Fig. 3 shows the results of analysis and simulated values for delay-bound violation probability as a function of delay bound. We observe that the analysis and simulation results agree reasonably well. As expected, when the number of users in system increases ($K=9$ vs. $K=5$), the packets in the buffer of the tagged user experience a higher delay-bound violation probability.

TABLE I
AMC transmission modes and fitting parameters for PER modeling.

| Mode | Modulation | Coding rate | Rate: $R_m$ | $a_m$ | $g_m$ |
|------|------------|-------------|-------------|--------|-------|
| 1 | BPSK | 1/2 | 0.5 | 1.4676 | 12.32 |
| 2 | BPSK | 3/4 | 0.75 | 0.7309 | 11.23 |
| 3 | QPSK | 1/2 | 1 | 0.3740 | 11.95 |
| 4 | QPSK | 3/4 | 1.5 | 0.1866 | 10.93 |
| 5 | 16-QAM | 1/2 | 2 | 0.0950 | 10.80 |
| 6 | 16-QAM | 3/4 | 3 | 0.0405 | 10.48 |
| 7 | 64-QAM | 2/3 | 4 | 0.0150 | 10.75 |

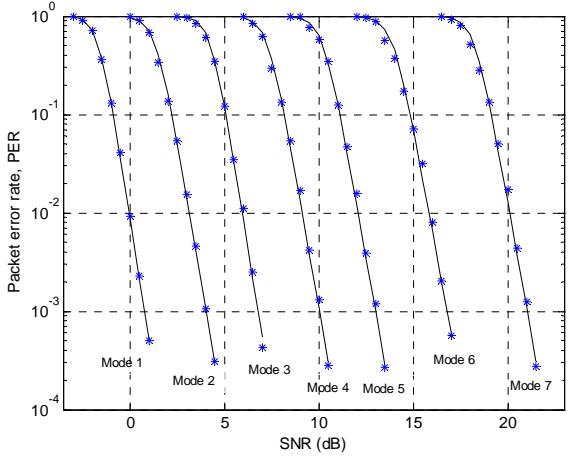

Figure 2. PER for AMC transmission modes in Table I: stars denote simulated PER, and solid lines indicate the fitted curves.

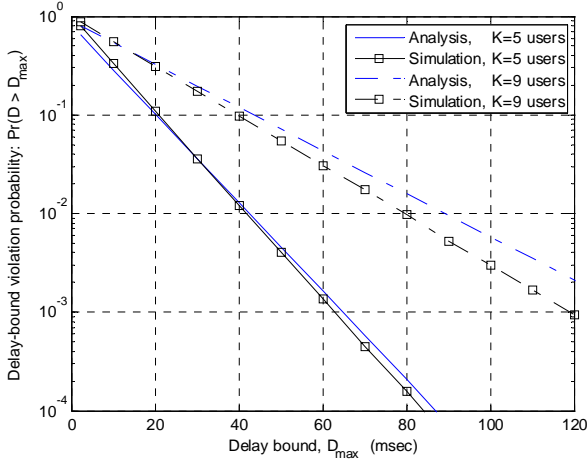

Figure 3. The analysis and simulation results for the delay-bound violation probability of the tagged user, $\rho_s$=49.1 kbits/s.

In order to illustrate the effect of arrival traffic statistics on the delay performance, in Fig. 4, we plot the delay-bound violation probability versus the mean ON-state period ($T_{ON}$) for three different average source rates. Note that for each group of curves in this figure, both $\rho_s$ and $v$ are fixed, therefore, the value of $T_{ON}$ reflects the burstiness of the arrival source. From Fig. 4, we see that the system with joint power adaptation and AMC provides a smaller delay probability in comparison with the CP-AMC scheme. In particular, for the reference source ($T_{ON}$ =10 frames), the AMC scheme with power control reduces the delay probability by 22%, 60%, and 66.2%, when compared to the CP-AMC scheme for the three assumed average source rates of $\rho_s$ =49.1, $\rho_s$=98.2, and $\rho_s$=147.3 kbits/s, respectively. We also observe that as the average source rate increases ($\rho_s$ =147.3 vs $\rho_s$=49.1), the burstiness of arrival traffic affects the delay violation probability much more strongly. In this case, power adaptation plays a more important role in reducing the delay violation probability.

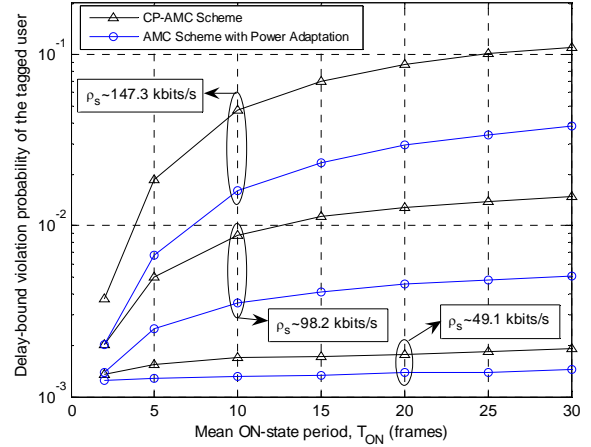

Figure 4. Delay probability vs. $T_{ON}$ for different average source rates, $K$=5 users (only simulation results are depicted).

## 7. APPENDIX 1

Here, we present a proof for Proposition 1. Since the r.v.s $x^{(1)}$ and $\tilde{x}$ are independent, we get the probability $P_j, j \geq 1$, as follows

$$\begin{aligned}
P_j &= \Pr\left(x^{(1)} \geq \tilde{x}, x^{(1)} \in [\xi_{j-1}, \xi_j]\right) \\
&= \int_0^{\xi_j} \Pr\left(x^{(1)} \geq x, x^{(1)} \in [\xi_{j-1}, \xi_j]\right) f_{\tilde{x}}(x)dx \\
&= \int_0^{\xi_{j-1}} \Pr\left(x^{(1)} \in [\xi_{j-1}, \xi_j]\right) f_{\tilde{x}}(x)dx \\
&\quad + \int_{\xi_{j-1}}^{\xi_j} \Pr\left(x^{(1)} \in [x, \xi_j]\right) f_{\tilde{x}}(x)dx \\
&= \int_{\xi_{j-1}}^{\xi_j} f_x(x)dx \cdot \int_0^{\xi_{j-1}} f_{\tilde{x}}(x)dx + \int_{x=\xi_{j-1}}^{\xi_j} f_{\tilde{x}}(x)\int_{y=x}^{\xi_j} f_x(y)dy\,dx
\end{aligned}$$
(23)

Also, the probability $P_0$ is given by

$$P_0 = 1 - \sum_{j=1}^{M} P_j = \int_{x=0}^{\infty} f_{x^*}(x) \int_{y=0}^{x} f_x(y)dy\,dx.$$
(24)

∎

## 8. APPENDIX 2

Here, we present a proof for Proposition 2. The probability $P_j, j \geq 1$, is given by

$$P_j = \Pr\left(J_n^{(1)} = J_n^* = j, V_n \leq 1/U_n\right)$$

$$= \sum_{i=1}^{K} \Pr\left(J_n^{(1)} = J_n^* = j, V_n \leq 1/U_n, U_n = i\right)$$

$$= \sum_{i=1}^{K} \Pr\left(V_n \leq 1/U_n \,\Big|\, J_n^{(1)} = J_n^* = j, U_n = i\right)$$
$$\times \Pr\left(J_n^{(1)} = J_n^* = j, U_n = i\right)$$

$$= \sum_{i=1}^{K} \Pr(V_n \leq 1/i) \cdot \Pr\left(J_n^{(1)} = j, J_n^{(2)} \leq j, \ldots, J_n^{(K)} \leq j, U_n = i\right)$$

$$= \sum_{i=1}^{K} \frac{1}{i}\binom{K-1}{i-1}\pi_j^i\left(\sum_{m=0}^{j-1}\pi_m\right)^{K-i} = \frac{1}{K}\sum_{i=1}^{K}\binom{K}{i}\pi_j^i\left(\sum_{m=0}^{j-1}\pi_m\right)^{K-i}$$

$$\stackrel{(a1)}{=} \frac{1}{K}\left[\left(\sum_{m=0}^{j}\pi_m\right)^K - \left(\sum_{m=0}^{j-1}\pi_m\right)^K\right]$$

$$= \frac{1}{K}\left[\left(\int_0^{\xi_{j+1}} f_x(x)dx\right)^K - \left(\int_0^{\xi_j} f_x(x)dx\right)^K\right]$$

$$= \frac{1}{K}\left([F_x(\xi_{j+1})]^K - [F_x(\xi_j)]^K\right) \quad (25)$$

where $\pi_m = \int_{\xi_m}^{\xi_{m+1}} f_x(x)dx$, and (a1) is derived using the following equality

$$(X + \pi_j)^K - X^K = \sum_{i=1}^{K}\binom{K}{i}\pi_j^i \cdot X^{K-i}. \quad (26)$$

∎